\documentclass{egpubl}
\usepackage[subtle]{savetrees}
\JournalSubmission    
\usepackage[T1]{fontenc}
\usepackage{dfadobe}  
\usepackage{url}
\usepackage{framed,multirow}
\usepackage[x11names]{xcolor}
\usepackage{enumitem}
\usepackage{lineno}
\UseRawInputEncoding

\usepackage{cite}  
\BibtexOrBiblatex
\electronicVersion
\PrintedOrElectronic
\ifpdf \usepackage[pdftex]{graphicx} \pdfcompresslevel=9
\else \usepackage[dvips]{graphicx} \fi

\usepackage{egweblnk} 
\usepackage{pdfpages}
\usepackage{graphicx}

\usepackage{xcolor}

\title[Style Brush: Guided Style Transfer for 3D Objects]%
      {Style Brush: Guided Style Transfer for 3D Objects}

\author[Kov{\'a}cs et al.]
{\parbox{\textwidth}{\centering {\'A}ron Samuel Kov{\'a}cs \orcid{0000-0002-0849-9032}, Pedro Hermosilla, and Renata G. Raidou \orcid{0000-0003-2468-0664}
        }
        \\
 {\parbox{\textwidth}{\centering TU Wien, Austria}
        }
}

\begin{document}

\teaser{
\vspace{-30pt}
\includegraphics[width=1\linewidth]{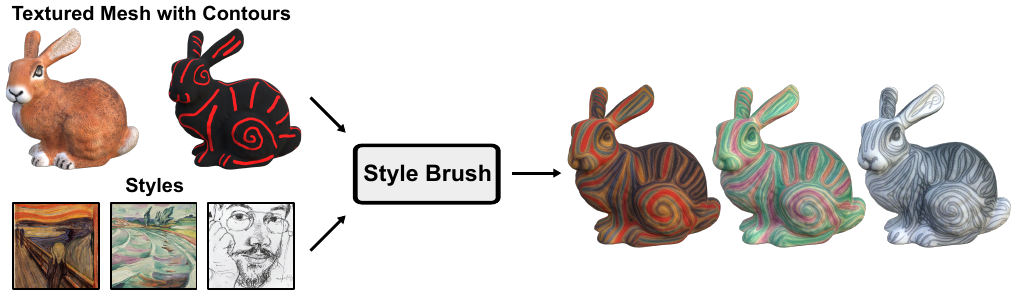}
\centering
 \caption{Style Brush takes a textured mesh, an additional texture with user-determined contours, and one (or multiple) style images as input. By optimizing the input texture and guiding the synthesized patterns with the contours, our method generates high-quality, stylized textures in just a few minutes that faithfully adhere to the directions specified by the user. Here, we demonstrate Style Brush using multiple style images, with each one producing a stylized texture; however, it also supports combinations of multiple styles, as we showcase later in our paper.}
\label{fig:teaser}
}

\maketitle
\begin{abstract}
We introduce Style Brush, a novel style transfer method for textured meshes designed to empower artists with fine-grained control over the stylization process.
Our approach extends traditional 3D style transfer methods by introducing a novel loss function that captures style directionality, supports multiple style images or portions thereof, and enables smooth transitions between styles in the synthesized texture.
The use of easily generated guiding textures streamlines user interaction, making our approach accessible to a broad audience.
Extensive evaluations with various meshes, style images, and contour shapes, demonstrate the flexibility of our method and showcase the visual appeal of the generated textures.


\end{abstract}  
\section{Introduction}

Style transfer refers to the process of applying a visual style of one image (e.g., a painting) to the content of another image, object, or scene---often by means of deep learning models~\cite{gatys2015styletransfer}. 
The image that provides artistic features, such as textures, colors, and/or brushstrokes, is referred to as \emph{style}. 
The style is applied to the \emph{content}, which is the counterpart that contributes to the structure. 
In this case, individual salient parts, whole objects, their arrangement, etc., are kept, but modified so that patterns from the style are used to construct them.

This process is especially interesting in a 3D context, as it is often necessary to create a large number of simple objects that share a visual style.
This is the case when game artists create backgrounds or environmental objects in a scene.
This process can be both tedious and time-consuming, making it an ideal target for automation.
Alternatively, style transfer could be used to investigate whether a visual style matches the artist's vision and suits their needs, before dedicating considerable resources to manually applying the style, possibly using the generated artifact as a starting point.

While relatively simple for images, using neural network-based approaches becomes much more complex for 3D objects and scenes.
In contrast to simple 2D images, it is necessary to ensure multi-view consistency, so that the patterns observed while moving in the scene stay coherent.
With the currently available technologies, building patterns that span across significant parts of a 3D object or scene requires using a relatively slow optimization-based process so that many viewpoints are in agreement as to which patterns should be used where~\cite{mordvintsev2018styletransferrender, gutierrez2019volumetexture, hollein2022styletransferrender, zhang2022arf, zhang2024coarf, kovacs2024gstyle}.

However, when utilizing such powerful tools, many of the creative decisions are left up to the more or less fully automatic random processes.
When artists use such a tool, they relinquish a lot of control, which may not be suitable for their projects.
For this reason, some methods allow the user to keep a certain degree of guidance to influence the creation process, e.g., by determining which patterns should be used where, the size of transferred patterns, or the directionality of transferred patterns~\cite{wu2019direction, reimann2022controlling}.
Unfortunately, to allow more complex guidance, e.g., determining directions of stroke-like patterns, existing methods either sacrifice quality or heavily constrain the type of possible interaction.
We are also not aware of any method that would allow these types of complex interactions specifically in 3D.

In this paper, we propose Style Brush, a novel artistic 3D style transfer method that allows the user to guide the stylization process by determining which style patterns should be used on which parts of the 3D objects, and the directionality of the synthesized patterns (Figure~\ref{fig:teaser}).
Our approach is based on using a differentiable renderer to achieve coherent stylization across different viewpoints, delivering a high-quality stylized mesh in just a few minutes.
For ease of use, we focus on using meshes, as creating \textit{guiding textures} for them is a straightforward task that potential users can handle. Even then, our approach requires only basic painting of guiding lines and regions.
In our evaluation, we show that our approach generates visually appealing mesh textures that respect the user-defined guidance, using a large variety of different textured meshes and styles.
We evaluate our method with different kinds of contour shapes (e.g., straight lines, spirals, circles) and style regions, showcasing our method's flexibility.
\section{Related Work}

Our work builds upon advances in style transfer, mainly for 3D objects and scenes. In this section, we provide an overview of neural style transfer methods for both 2D and 3D cases.

\noindent \textbf{2D Style Transfer.} Gatys et al.~\cite{gatys2015styletransfer} proposed a neural style transfer method, which uses the style of an image and the content of another image, to synthesize an image that uses patterns from the style image to construct the content of the other image.
This method iteratively optimizes the content image to match the style statistics, expressed as Gram matrices of hidden layers of a neural network, that was originally trained for classifying images.
At the same time, their approach also tries to maintain the original content, by matching the outputs of the network's layers as applied to the original content image.
Later, Gatys et al.~\cite{gatys2015texturesynthesis} modified their approach to create new textures based on the provided style image.
They accomplish this by using a noise image as the starting point and not trying to keep the original content.

The approach of Gatys et al. served as an important starting point and inspired many follow-up papers.
These papers can be roughly categorized into two main groups.
The first category of approaches is optimization-based.
They optimize the initial image using a variety of different losses to better preserve certain aspects of the style image, possibly using different neural architectures.
These include transferring features on multiple scales~\cite{gu2018arbitrarystyleloss}, utilizing Generative Adversarial Networks (GANs)~\cite{jetchev2017texturesynthesisgan}, or diffusion models~\cite{zhang2023diffusionstyletransfer, wang2023stylediffusion, chung2024style}.
While some directly base their losses on the original approach, i.e., matching Gram matrices between images, other works search for nearest neighbors in the feature space to minimize the distance between them~\cite{kolkin2019styleoptimaltransportlossnn, chen2016styletransferfastnn, liao2017imageanalogylossnn, li2016combiningmarkovnn, zhang2022arf, zhang2024coarf}.
Furthermore, different types of networks and losses can transfer different kinds of patterns and thus using multiple of them at the same time can complement each other~\cite{kovacs2024gstyle}.
The other group of approaches explicitly minimizes the loss function in a single feed-forward pass~\cite{chen2016styletransferfastnn, huang2017arbitrarytransferff, an2021artflowff}. They are generally faster than the optimization-based approaches; however, some may require lengthy pre-training for each style and their results tend to be of lower quality than the optimization-based approaches.

\noindent \textbf{3D Style Transfer.} 3D style transfer methods aim to modify the appearance of 3D objects or scenes to match the appearance of a given style.
Visual style transfer is, in a way, inherently 2D, as human vision is based on image projection.
This implies that if they are sufficiently adapted, 2D methods can be used for 3D style transfer.
To this end, some methods utilize a differentiable renderer to create 2D images~\cite{mordvintsev2018styletransferrender, zhang2022arf, zhang2024coarf, kovacs2024gstyle}, other methods slice the 3D volume~\cite{henzler2020neuraltexture, gutierrez2019volumetexture, zhao2022stsgan, chen2010high, kopf2007solid}, or directly work with the 2D surface manifold if the 3D representation is surface-based~\cite{kovacs2024surface}.

Depending on the chosen 3D representation, different methods have to deal with the inherent limitations of the representations.
The source of the 3D objects and scenes often influences the choice of these representations.
Meshes are in many cases the industry standard and, as such, are also utilized as the underlying representation~\cite{mordvintsev2018styletransferrender, hollein2022styletransferrender, kovacs2024surface}.
However, if the 3D objects are obtained from real-world measurements, other representations are often utilized.
Cao et al.~\cite{cao2020styletransferpointclouds} use point clouds, this representation, however, contains holes and render-based approaches would need to robustly deal with them as the background may be visible even through solid surfaces.
Lately, learned volumetric representations have become popular~\cite{mildenhall2020nerf, chen2022tensorf, kerbl3dgaussians}, mainly owing to their high reconstruction quality of real-world scenes.

Similarly to the 2D methods, 3D methods can be divided into the same two categories, either iterative or using a single feed-forward pass.
A significant part of 3D style transfer for a style that contains large patterns or features is to embed the features into the chosen 3D representation, such that they are consistent across different viewpoints and still true to the style.
However, single-pass methods are currently unable to do so and mostly focus on changes that do not require synthesizing large-scale patterns, mainly matching color statistics or relighting~\cite{liu2023instant, liu2023stylerf, xu2024styledyrf}.

\begin{figure*}[htp]
  \centering
  \includegraphics[width=\linewidth]{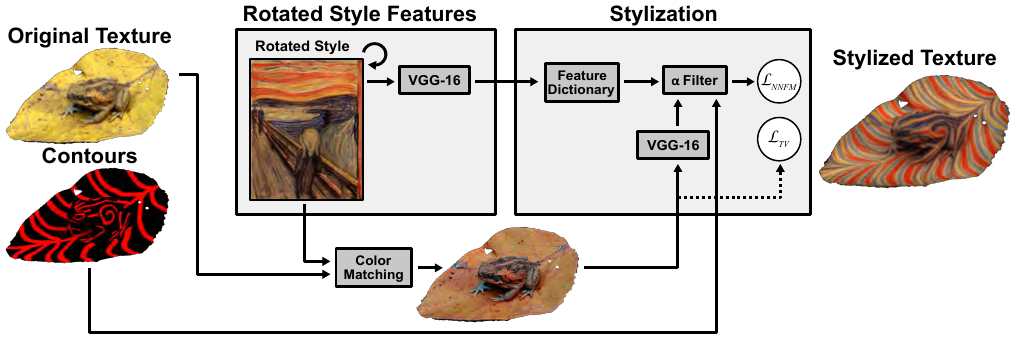}
  \caption{\textbf{Overview of our method: } We take a textured mesh, an additional texture with guiding lines, and a style image. Initially, we extract rotated style features by rotating the style image, passing it through a feature extractor, and assigning to each feature its directionality, thus building an angle-based feature dictionary. Next, we perform a color matching step between the style image and the original texture. Finally, we optimize the texture by rendering the mesh from multiple viewpoints, extracting features, and calculating the style loss by matching the rotated style features to the directions defined by the rendered contours. We minimize the total variation loss to suppress noise and repeat the entire process until convergence. }
  \label{fig:overview}
\end{figure*}

To create these large-scale patterns, current methods rely on relatively slow optimization processes, where the patterns are progressively built or dissolved to reach consensus across many different viewpoints or slices.
Mordvintsev et al.~\cite{mordvintsev2018styletransferrender} use a differentiable renderer in a very straightforward way, showcasing that simply rendering a 3D object and then applying a simple style loss produces textures of similar quality to 2D images, but in 3D.
H{\"o}llein et al.~\cite{hollein2022styletransferrender} extend this approach to scenes by utilizing a depth and angle-aware optimization, accounting for surfaces that may not be aligned to the camera's viewing plane, and that objects may have different screen space sizes depending on the distance from the camera.
Gutierrez et al.~\cite{gutierrez2019volumetexture} take a different approach and create a volume texture that is stylized by applying a 2D style loss to slices of the volume texture, however, this approach is not aware of any 3D object that may be sculpted from the volume and is thus less sufficient for more complex tasks.
With the advent of learned 3D representations, many methods extend 3D-based style transfer methods to these representations, oftentimes having to account for the special properties of these representations compared to simpler meshes or volumes~\cite{zhang2022arf, zhang2024coarf, kovacs2024gstyle}.

\noindent \textbf{Guidance.}
Style transfer guidance means that the user can meaningfully influence the stylization process.
This can take many forms. In the context of artistic style transfer, the simplest is selecting which style images to use and optionally applying simple transformations such as cropping or masking to select specific patterns. This is used in the work of Zhang et al.~\cite{zhang2024coarf}, to stylize 3D scenes based on semantic-similarity maps.
Another possibility is to use semantic label maps.
By using the semantic layout as input and stacking convolutional, normalization, and nonlinear layers, it is possible to synthesize realistic-looking images, with various features, such as "rock", "tree", etc., at user-defined locations ~\cite{isola2017image, wang2018high}.
This can be further enhanced by including a spatially adaptive, learned transformation to modulate activations~\cite{park2019semantic}.
Another form of guidance is to direct features, so that the user can define the direction of brush strokes, pen strokes, etc.
This can be accomplished by training a neural network to detect the directions and building a loss function with it to align the flow of patterns with the user-defined flow~\cite{wu2019direction}.
Another possibility is to use reversible content transformations to adjust the orientation of directed patterns, which can also be used for other transformations such as scaling~\cite{reimann2022controlling}.

In our current work, we use meshes as the underlying representation because interacting with them is easier than the learned volumetric representations, like NeRFs or Gaussian Splatting.
As we aim to synthesize patterns that span a significant portion of the 3D objects or scenes, our approach needs to be optimization-based. Thus, we utilize a loss based on finding nearest neighbors in feature space, due to the high-quality results that this type of loss can accomplish~\cite{zhang2022arf, zhang2024coarf, kovacs2024gstyle}.
Furthermore, we want to empower users to interact with the stylization process, as artistic styles often feature directed patterns utilizing brush strokes.
To do so, we allow brush-like interaction by creating textures that indicate the preferred directions of patterns on the meshes at hand. 
These textures allow the user to select which styles (or parts thereof) should be applied on dedicated parts of the mesh.
To our knowledge, our method is the first one to allow this kind of guided interaction-based style transfer for 3D objects, allowing the users to get high-quality textures that better correspond to what they want, instead of getting a collection of random patterns based on under-constrained hallucinations.
\section{Methodology}

In this section, we describe our proposed algorithm for the stylization of meshes with the use of sketch-like guidance. Firstly, we provide an overview of our approach, which is followed by a detailed explanation of each step.

\subsection{Overview}

A schematic overview of Style Brush is provided in Figure~\ref{fig:overview}.
Our algorithm takes as input a \textit{mesh} with the original content $\mathcal{T_C}$, a set of $n$ \textit{style images} $\mathcal{I}^{n}_{\mathcal{S}}$, a \textit{directional guidance texture} $\mathcal{T_D}$, and possibly a \textit{style mask texture} $\mathcal{T_S}$.
The \textit{input mesh} may already have an initial texture (e.g., colors, patterns, etc.), but our algorithm can also work without it. 
The \textit{set of style images} are $n$ 2D images that define one or more artistic styles to be applied to different regions of the mesh, such as painting styles, patterns, or textures that influence the final look of the textured mesh.
The \textit{directional guidance texture} provides directional information for applying the style (Figure~\ref{fig:overview}, "Contours").
It can be used to control the flow of brush strokes or align patterns along a specific direction across the surface of the mesh.
The \textit{style mask texture} acts as a mask or segmentation map, indicating where the different style images should be applied to specific parts of the 3D mesh.
Please note that Figure~\ref{fig:overview} shows an example with only one style image and without style masks.

Our method renders multiple views of a 3D mesh and then iteratively modifies its texture with a differentiable renderer.
To enforce the desired guidance, we ensure that the synthesized texture follows the direction indicated by the directional guidance texture.
To do this, we first compute the edge tangent flow~\cite{kang2007coherent} of each style image, extract features with a neural extractor, and assign to each feature its direction based on the flow (Figure~\ref{fig:overview}, "Rotated Style Features").
Before optimization, we match the color statistics of the texture $\mathcal{T_C}$ to the style image, which speeds up convergence and ensures accurate color reproduction (Figure~\ref{fig:overview}, "Color Matching").
Finally, we optimize the texture by applying a style loss that selects and minimizes the distance between the extracted features of the rendered texture and the style image, taking into account the guidance textures $\mathcal{T_D}$ and $\mathcal{T_S}$.
Here, we also minimize the total variation of the rendered images to suppress noise (Figure~\ref{fig:overview}, "Stylization").
The entire optimization process runs until convergence, ensuring that the final stylized texture faithfully reflects the desired artistic characteristics.
Detailed descriptions of each of these steps are provided in the following subsections.

\subsection{Dictionary of Rotated Style Features}

Artistic style images tend to contain directional patterns, e.g., brush strokes, pencil strokes, pen strokes, etc.
When applying these styles to a 3D mesh, the patterns must align with the directional guidance provided by the user.
In our case, we determine the directionality of patterns by computing the edge tangent flow, an approach also utilized by Wu et al.~\cite{wu2019direction}.
Edge tangent flow (ETF) is a smooth direction field that corresponds to the flow of patterns in the input image. 
In our implementation, we utilize the ETF algorithm by Kang et al.~\cite{kang2007coherent}.
This algorithm is based on computing the gradient of an image and subsequently iteratively smoothing it, accounting for the gradient magnitude.
Note that, in general, patterns flowing in a certain direction are indistinguishable from patterns going in the exact opposite direction.
Thus, we treat them equally, meaning that we only need to consider orientations in the range of $[0, \pi)$ radians rather than $[0, 2\pi)$. 
This is also the case for the ETF algorithm, which outputs an image $\mathcal{I_{ETF}}$ where each pixel is in $[0, \pi)$ radians, indicating the flow of features.
The ETF algorithm is parametrized by the kernel radius used during smoothing and the number of iterations.
The choice of these parameters impacts the extent to which small patterns can be detected, particularly when they flow in a different direction from their surrounding region~\cite{kang2007coherent}.
We set both of these parameters to 10.

Having computed $\mathcal{I_{ETF}}$ for a particular style image, we discretize the directions into angle sets. 
Each angle set $\mathcal{S_\alpha}$ contains pixels having a similar direction $\alpha$ within a parametrized tolerance $\tau$. 
We represent these angle sets as an integer image $\mathcal{I_\alpha}$.
This discretization step helps us to organize the directionality of the patterns in a way that facilitates accurate matching, pairing, and application of patterns in the style transfer process.
We set $\tau$ to $5^\circ$, meaning that the first set contains pixels with ETF of $[177.5^\circ, 180^\circ) \cup [0^\circ, 2.5^\circ)$, the second $[2.5^\circ, 7.5^\circ)$, and so on.

Once we have computed the angle set image $\mathcal{I_\alpha}$, we generate rotated versions of the style image and its $I_\alpha$ in increments of $\tau$ and process each with a pre-trained VGG-16 network.
Having done so, we extract features from the VGG-16's~\cite{simonyan2014vgg16} hidden layers.
Inspired by Kolkin et al.~\cite{kolkin2022nnst}, we extract features from the first seven layers and resize the feature maps to $\frac{W}{4}\frac{H}{4}$, where $W$ and $H$ are the width and height of the input image.
This choice is further motivated by the fact that we need to store features for every rotated version of the style image, which is very demanding on the available GPU memory.
This number can be further adjusted, thus having fewer or more style features, which may impact the stylization quality.
Using the rotated $I_\alpha$, we extract the features associated with each direction $\alpha$ and store them in feature sets $F^S_\alpha$.

\subsection{Stylization}
\label{ssec:stylization}
Once we have prepared the style features and masks, we can start the stylization process.
Our approach is render-based, which means we need to render the mesh from many different viewpoints, applying our style loss, and back-propagating the gradient to the mesh's texture.
The object's texture can be either a pre-made one, $\mathcal{T_C}$, in case it is desired that the resulting texture keeps its features, or a randomly initialized texture, in case there is no available texture or the original content is not important.
For simplicity, we will refer to both as $\mathcal{T_C}$.
As our primary goal is to stylize single objects, we assume that a uniform distribution of viewpoints on a sphere around the object sufficiently covers all parts of the mesh.
If that is not the case, an extension of our method may consider a different distribution of viewpoints to ensure better coverage of the object.
We generate a uniform set of viewpoints using the Fibonacci sphere, rendering the mesh from the points on the sphere pointing towards the center of the mesh.

For each viewpoint, we compute which directional features to use for which part of the rendered image.
Again, we utilize Kang et al.'s ETF algorithm~\cite{kang2007coherent}.
However, instead of computing the ETF of the rendered image, we allow the user to guide the ETF computation by defining a texture $\mathcal{T_D}$, which contains guiding lines, as depicted in Figure~\ref{fig:overview} (see "Contours").
We compute the ETF in screen space.
First, we render the mesh with the contour texture.
In order to have a non-zero ETF everywhere, we detect the edges of the rendered contours and compute the distance to the nearest edge pixel, and we use this distance image to compute the ETF.
We set the kernel radius and the number of iterations to 5, and then we discretize it into $\mathcal{R_\alpha}$ using the same $\tau$ from the style feature extraction process.
Because $\mathcal{R_\alpha}$ and the directional feature sets $F^S_\alpha$ share the same discretization granularity, we can directly match and select which style features to apply to specific regions of the stylized image.

Our style loss is based on the work of Zhang et al.~\cite{zhang2022arf} and Kolkin et al.~\cite{kolkin2022nnst}.
Both use style losses based on nearest neighbor feature matching (NNFM), utilizing a pre-trained VGG-16 network as the feature extractor to capture high-frequency style patterns.
Unlike the widely used loss of Gatys et al.~\cite{gatys2015styletransfer}, which uses Gram matrices of features to define the distance between styles, the NNFM loss finds the nearest neighbors in feature space and minimizes the distance between them.
Let $\mathcal{F^R}$ and $\mathcal{F^S}$ be the extracted feature maps of a rendered image $\mathcal{R_C}$ of the mesh using the $\mathcal{T_C}$ texture and a style image respectively, and let $\mathcal{F}(i)$ be the $i$-th feature from the map.
Then, the NNFM loss is defined by Zhang et al. as:
\begin{equation}
    \mathcal{L}_{NNFM}(\mathcal{F}^{R}, \mathcal{F}^{S}) = \frac{1}{N}\sum_{i}\min_{j} \phantom{i}D(\mathcal{F}^{R}(i), \mathcal{F}^{S}(j))
\end{equation}
where $N$ is the number of features in $\mathcal{F^R}$ and $D$ is the cosine distance between vectors.

Kolkin et al. make further adjustments to the loss by finding the nearest neighbors for each \textit{layer} separately, then combining them into a mixed feature vector, and aligning the means of the features, such that they are centered at 0.
This allows the loss to synthesize more varied features and more effectively transfer patterns, depending on how well the style of the reference image aligns with the style of the optimized content image.
Let $\mathcal{F}(L, i)$ be the $i$-th feature in the $L$-th layer of the feature maps.
Then, the adjusted NNFM loss is defined as:
\begin{equation}
    \mathcal{L}_{NNFM}(\mathcal{F}^{R}, \mathcal{F}^{S}) = \frac{1}{N}\sum_{i}\min_{j} \phantom{i}D(\bigcup_L(\mathcal{F}^{R}(L, i) - \mu^R_L, \mathcal{F}^{S}(L, j) - \mu^S_L))
\end{equation}
where $\mu^R_L$ and $\mu^S_L$ are the means of the $L$-th layers of $\mathcal{F}^{R}$ and $\mathcal{F}^{S}$.

We make further adjustments to this loss by placing features into sets based on their directionality and only finding the nearest neighbors in the sets that have the same direction.
Thus, our new style loss becomes:
\begin{equation}
    \mathcal{L}_{NNFM}(\mathcal{F}^{R}, \mathcal{F}^{S}) = \frac{1}{N}\sum_\alpha\sum_{i}\min_{j} \phantom{i}D(\bigcup_L(\mathcal{F}^{R}_\alpha(L, i) - \mu^R_L, \mathcal{F}^{S}_\alpha(L, j) - \mu^S_L))
\end{equation}
where $\mathcal{F}^{R}_\alpha$ was generated from $\mathcal{F}^{R}$ using $\mathcal{R_\alpha}$ with the same process that we used for style images without additional rotations.

Furthermore, we use the total variation loss $\mathcal{L}_{TV}$ to reduce noise.
Note that we only compute total variation with non-background pixels to avoid bleeding of the background into the mesh.
Thus, our final loss is defined as:

\begin{equation}
    \mathcal{L} = \mathcal{}\mathcal{L}_{NNFM} + \lambda\mathcal{L}_{TV}
\end{equation}

\noindent where $\lambda$ is the weighting coefficient for the total variation loss.

\subsection{Style Color Matching}

The NNFM loss exhibits the ability to synthesize high-level patterns, however, it struggles with correctly transferring color between the style image and the optimized texture.
This can be explained by the fact that features of hidden layers do not necessarily carry all color information.
The color of the converged texture $\mathcal{T_C}$ heavily depends on its initial stage. 
For this reason, we match the distribution of colors between the texture and the style image, before we start the optimization discussed in Section~\ref{ssec:stylization}.

Let $C$ be the matrix $R^{N \times 3}$ of the used texels of $\mathcal{T_C}$, i.e., the parts of the texture that may be potentially sampled during rendering, and let $S$ be the matrix $R^{M \times 3}$ of the style image's pixels.
As in the work of Zhang et al.~\cite{zhang2022arf}, we solve analytically for a linear transformation $M$, so that $E(MC)=E(S)$ and $Cov(MC)=Cov(S)$.
We show the effectiveness of this step in our ablation studies (see Section~\ref{sec:ablation}).

\subsection{Multiscale Feature Transfer}

Images often contain patterns at multiple scales, e.g., ranging from fine brush strokes to larger structures composed of those strokes.
To effectively transfer patterns at different scales, we employ a simple yet effective approach inspired by Kolkin et al.~\cite{kolkin2022nnst}.
First, we downscale the texture $\mathcal{T_C}$, the style images $\mathcal{I}^{n}_{\mathcal{S}}$, and render the meshes at a reduced resolution, all using the same downsampling ratio.
Next, we perform style transfer, ensuring that only features recognizable by the feature extractor at the given scale are transferred.
Finally, we upsample to match the next downsampling ratio, which is defined relative to the original sizes.
We repeat this process iteratively until we match the original resolution.
In our results, we downsample once by 2, although the number of intermediate steps is parametrizable by the user.

After each intermediate step, we blend the original texture at that scale with the newly generated texture using a blending coefficient $\beta$.
This allows us to control the influence of larger patterns in the final result while preserving the original content's details, thus, it can also be thought of as a content weight.
Note that this blending step is done with the color-matched version of the content texture to ensure a suitable distribution of colors.
With this, we aim to encourage the formation of larger patterns.

\begin{figure*}[htp]
  \centering
  \includegraphics[width=\linewidth]{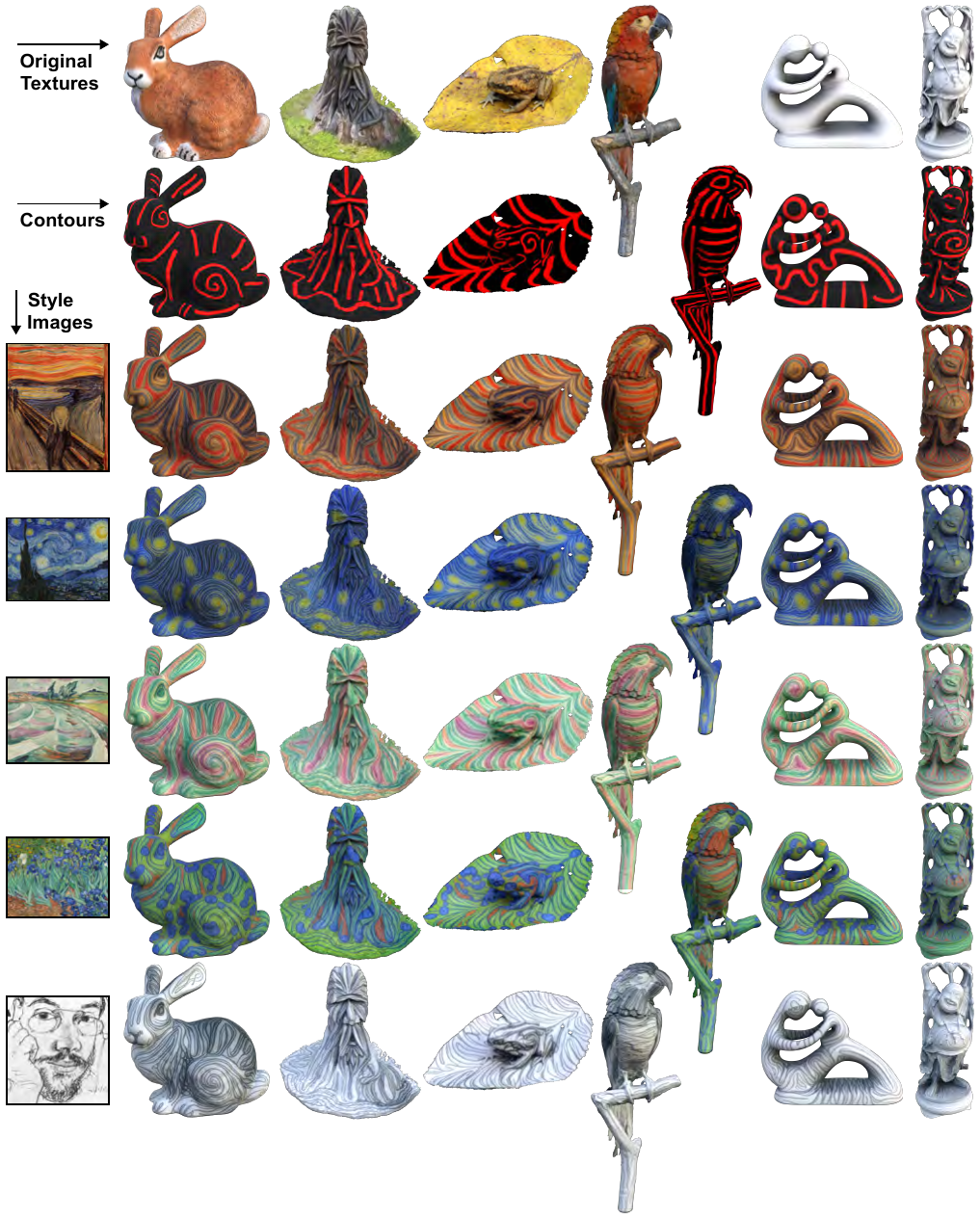}
  \caption{Results generated with Style Brush for six textured meshes (columns), using as input the initial textured mesh (first row), directional guidance in the form of contours (second row), and five style exemplars (last five rows).}
  \label{fig:main_results}
\end{figure*}

\subsection{Partial and Multiple Styles}

Furthermore, we also enable the use of partial style images (i.e., only specific regions of the style image are applied) and multiple style images (i.e., applying different style sources to distinct parts of the object) through the use of style masks. We outline the adjustments made to our method to support these two cases.

\noindent \textbf{Partial Styles.}
In case it is desired to only apply a part of the style image, the most straightforward way is to crop the style image.
However, this confines the selection to a simple rectangular region.
To allow for more control and flexibility, the user can create a binary style mask, which we apply during the creation of the rotated feature dictionary.
By rotating the mask, we retain only the features within the masked area.
The color matching step is only performed with the pixels from the active region.

\noindent \textbf{Multiple Styles.}
To support multiple styles, we utilize the style mask texture $\mathcal{T_S}$.
We render this texture during stylization to determine which style image features should be considered in the nearest neighbor search when computing the style loss.
Furthermore, $\mathcal{T_S}$ is used to determine which parts of the texture are used for which style image during the color matching step.

\begin{figure}[!b]
    \centering
      \includegraphics[width=1\linewidth]{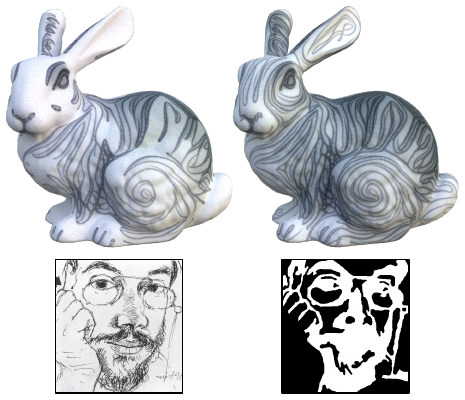}
    \caption{Two examples of our method: the one on the left uses the full style image, while the one on the right applies only the features corresponding to a mask (bottom right).}
    \label{fig:partial_style}
\end{figure}

\begin{figure*}[htp]
  \centering
  \includegraphics[width=\linewidth]{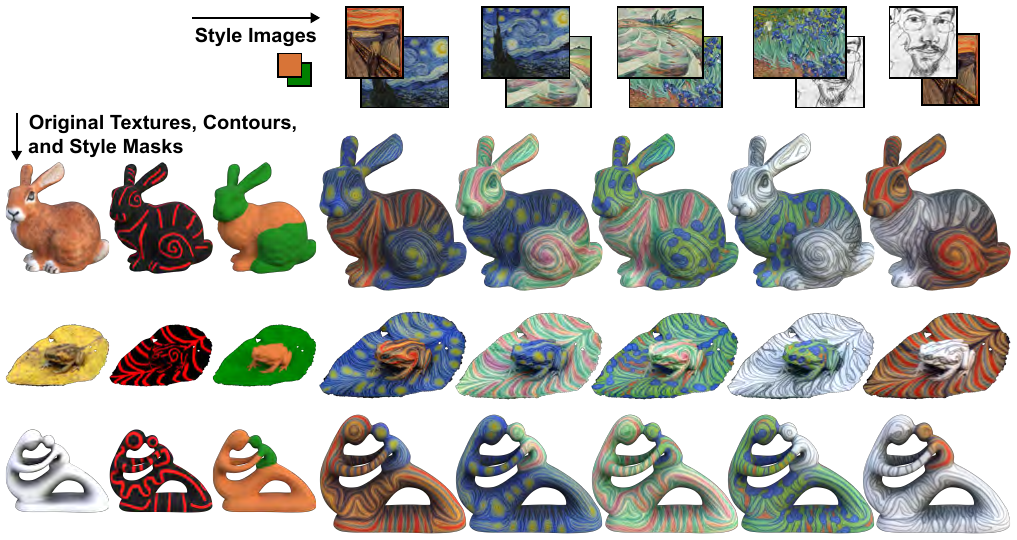}
  \caption{ The results of our method for three meshes and five different styles, where two styles are combined. Here, two different style images are used on different parts of the mesh, defined by the style mask texture (indicated with orange and green). The style mask texture is given as an input to Style Brush, in addition to the original textured mesh and the guiding contours. }
  \label{fig:multiple_styles}
\end{figure*}

\begin{figure}[!t]
    \centering
      \includegraphics[width=.75\linewidth]{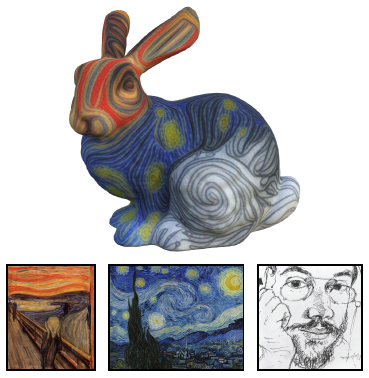}
    \caption{A stylized texture generated with our method, combining three style regions, each with a different style image.}
    \label{fig:three_styles}
\end{figure}

\subsection{Implementation Details}

We set the render resolution to $512^2$, unless the object is elongated, in which case we set it to $1024^2$.
We do this to ensure that the number of non-background pixels is approximately the same, so we achieve a similar level of detail for objects that are more round and for objects that are elongated (see Figure~\ref{fig:main_results} and compare the Stanford bunny and the macaw).
Furthermore, all of the textures we use are $2048^2$.

Note that rendering the geometry and the computation of ETF are computationally expensive.
For this reason, we precompute fragments, i.e., we associate pixels with texture coordinates, sample the guidance textures with the fragments, and compute the ETF.
We use the precomputed fragments during the stylization.
As our implementation of the ETF algorithm is CPU-based, computing many directional fields from multiple viewpoints is trivially parallelizable.
However, this comes at the cost of the memory needed to store this data.
To account for this, in our experiments, we generate 250 viewpoints around an object using the Fibonacci sphere.
For the optimization, we set $\lambda$ to $2\mathrm{e}{-5}$ and we use the Adam optimizer with the learning rate set to 0.01 and optimize for 1000 iterations for each multiscale step.
\section{Results}

In this section, we present an analysis of the results generated using our method.
We implemented our method in Python and Rust, utilizing PyTorch and PyTorch3D.
Our code will be made publicly available upon our paper's acceptance.

\subsection{Dataset}

We evaluated our method on a variety of publicly available meshes.
The meshes used for the evaluation are the following: the Stanford bunny and Happy Buddha from \href{http://graphics.stanford.edu/data/3Dscanrep}{\textcolor{blue}{the Stanford 3D scanning repository}}, \href{https://sketchfab.com/3d-models/the-green-man-druid-hill-park-c104c5e505ff4ad59f16a9df2c385559}{the Green Man} mesh by the user ``gerg'' licensed under \href{https://creativecommons.org/licenses/by/4.0/}{\textcolor{blue}{CC BY 4.0}}, a \href{https://sketchfab.com/3d-models/model-6-marine-toad-on-leaf-d3951b2882a140219a7fc9b61e0183a5}{marine toad on a leaf} mesh by DigitalLife3D licensed under \href{https://creativecommons.org/licenses/by-nc/4.0/}{\textcolor{blue}{CC BY-NC 4.0}}, a \href{https://sketchfab.com/3d-models/cuban-macaw-nhmw-zoo-vs-50796-52348a29605a4136a0e807942d3c12a6}{cuban macaw} mesh provided by the Natural History Museum Vienna licensed under \href{https://creativecommons.org/licenses/by-nc/4.0/}{\textcolor{blue}{CC BY-NC 4.0}}, and the \href{https://www.thingiverse.com/thing:456430}{Mother and Child} mesh by the user ``edcorusa'' licensed under \href{https://creativecommons.org/licenses/by-sa/3.0/}{\textcolor{blue}{CC BY-SA 3.0}}.
The models that were generated by scanning real-world objects were cleaned as appropriate to regularize their geometry.
The Stanford bunny, Happy Buddha, and the Mother and Child meshes are not textured.
We used a texture for the Stanford bunny taken from \href{https://web.archive.org/web/20220714184311/http://alice.loria.fr/index.php/software/7-data/37-unwrapped-meshes.html}{\textcolor{blue}{http://alice.loria.fr/index.php/software/7-data/37-unwrapped-meshes.html}} (the link leads to \href{https://archive.org/}{\textcolor{blue}{https://archive.org/}} as the website no longer exists), and the textures for the Happy Buddha and Mother and Child meshes were derived from their ambient occlusion textures. 
The other meshes already have a texture. 

Furthermore, we used five artistic style images that consist of directed patterns, i.e., brush strokes and pencil strokes: \textit{The Scream} and \textit{The Waves} by \textit{Edvard Munch}, \textit{The Starry Night} and \textit{Irises} by \textit{Vincent van Gogh}, and a self-portrait by \textit{Alexandr Benois} with a mask shown in Figure~\ref{fig:partial_style}. These have been chosen to reflect a diversity in patterns, colors, strokes, and other visual features. 

\subsection{Main Results}
Using the aforementioned meshes and styles we obtain the results presented in Figure~\ref{fig:main_results}, showcasing the flexibility of our approach across diverse styles. For example, the swirling brushstrokes of \textit{The Starry Night} (Figure~\ref{fig:main_results}, second row) are effectively used to stylize even meshes with intricate surface details such as the Green Man (Figure~\ref{fig:main_results}, second column) or the Happy Buddha (Figure~\ref{fig:main_results}, last column).  Also, the more structured pencil strokes of \textit{The Scream} (Figure~\ref{fig:main_results}, first row) or the \textit{Irises} (Figure~\ref{fig:main_results}, fourth row) stylize our meshes following the given guidance, proving the versatility of our method in handling different artistic elements.

\begin{figure*}[htp]
  \centering
  \includegraphics[width=\linewidth]{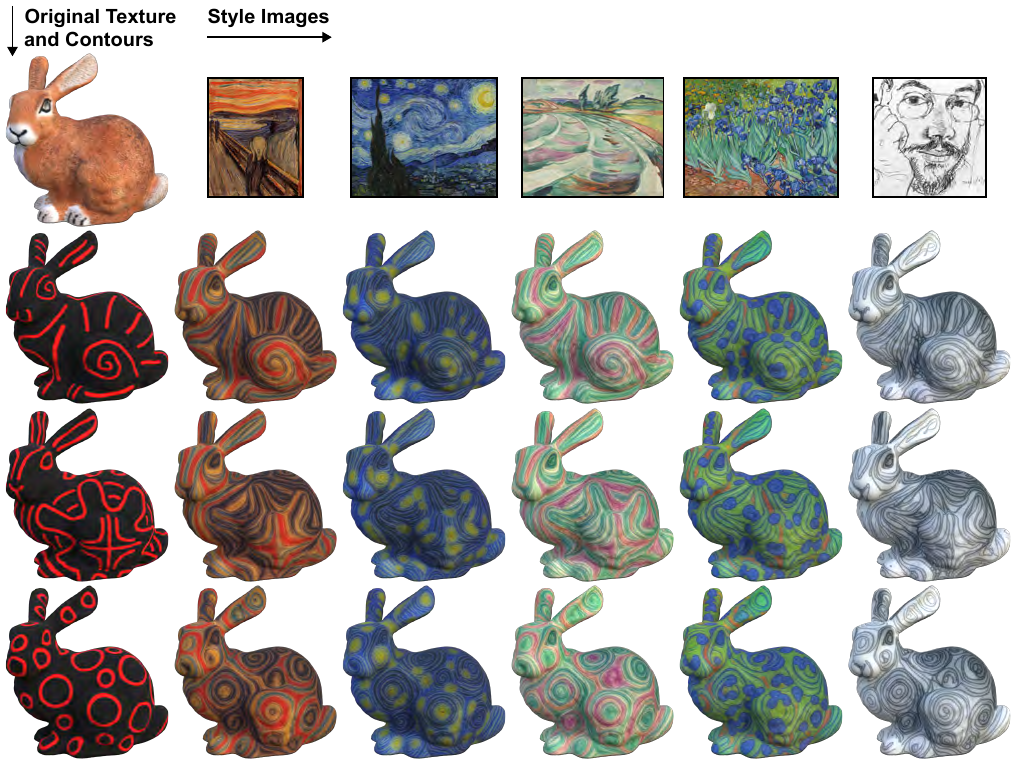}
  \caption{Results generated with our approach using three different directional guidance textures.}
  \label{fig:alternative_contours}
\end{figure*}

In Figure~\ref{fig:partial_style}, we show how our method can be fine-tuned to apply only a portion of the style, allowing for partial adaptation and offering fine control over the final appearance of the stylized mesh. This flexibility is particularly useful when specific stylistic elements need to be emphasized. Moreover, our approach can seamlessly handle multiple styles applied to different regions of the mesh (see Figure~\ref{fig:multiple_styles} and Figure~\ref{fig:three_styles}). This demonstrates how well our method adapts to complex and varied artistic inputs.
By altering the directional guidance textures, we enable further customization, allowing for a high degree of control over how each style is represented, as seen in Figure~\ref{fig:alternative_contours}.

In all those examples, we see that our method works effectively across different mesh topologies, handling both simple and complex shapes, making it a versatile tool for diverse applications. Overall, the combination of flexible style handling, control over directional guidance, and adaptability to different mesh structures makes our method highly effective in producing high-quality, stylized textures.

\begin{figure*}[htp]
  \centering
  \includegraphics[width=\linewidth]{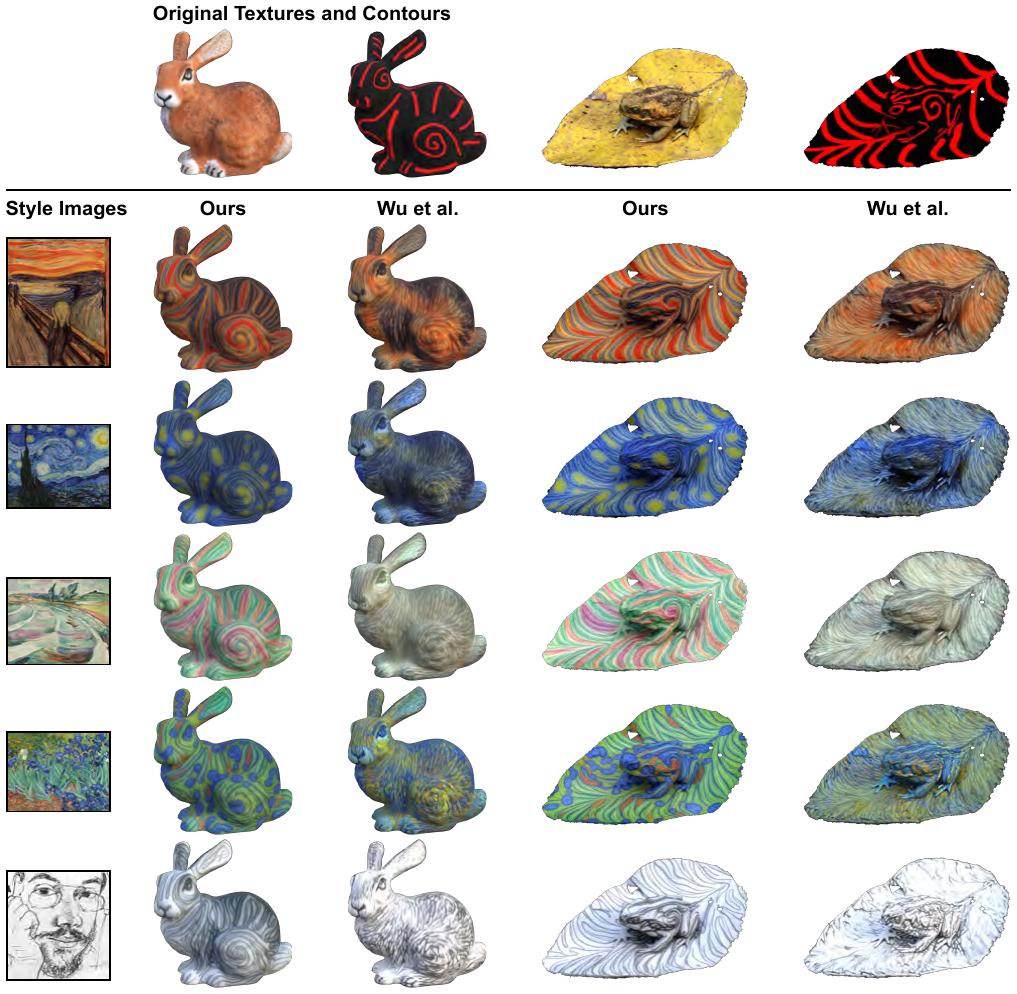}
  \caption{Results generated with our approach compared to those generated by the adapted (from 2D to 3D) method of Wu et al.~\cite{wu2019direction} for two textured meshes (the Stanford Bunny and the marine toad on a leaf mesh) and five style images.}
  \label{fig:comparison}
\end{figure*}

\subsection{Comparison to the State of the Art}

To the best of our knowledge, there is no published research on guided artistic style transfer for 3D objects.
As such, we are only able to compare our method with the work of Wu et al.~\cite{wu2019direction}, which is an optimization-based method for guided artistic style transfer in 2D.

Wu et al.'s approach is based on utilizing a differentiable ETF estimator and then using its estimate to compute an additional loss term, the mean squared difference between the desired directional field and the directional field of the currently optimized image.
Other than this, it uses Gatys et al.'s~\cite{gatys2015styletransfer} style and content losses.
As their code implementation is not public, we extended the implementation of Gatys et al. with Wu et al.'s directional loss utilizing a direct computation of ETF, which is differentiable, and employing the same differentiable renderer setup as we do for our method.
In this way, we aim to reproduce their results and extend them to textured meshes.

\noindent \textbf{Qualitative Comparison.}
A comparison between our method and the method of Wu et al.~\cite{wu2019direction} can be seen in Figure~\ref{fig:comparison}.
We compare these two methods using two meshes (the Stanford bunny and the marine toad on a leaf) and five style images.

We observe that even though the method of Wu et al. is capable of producing patterns partially resembling the style images, their generated textures are not as truthful to the style images as ours. For instance, notice the difference in the generated stars when using \textit{The Starry Night} (Figure~\ref{fig:comparison}, second row) or the colorful brush strokes when using \textit{The Wave} (Figure~\ref{fig:comparison}, third row).
All of the generated textures using Wu et al.'s approach are densely covered with lines and curves aiming to resemble brush or pencil strokes, though the patterns themselves fail to capture the visual quality of the brush or pencil strokes.
For instance, notice the case of \textit{The Scream} style (Figure~\ref{fig:comparison}, first row) or the \textit{Irises} style (Figure~\ref{fig:comparison}, fourth row), where Style Brush excels in comparison to Wu et al.'s approach. 

While we observe that the directions of the patterns generated with Wu et al.'s approach correspond to the guiding lines to an extent, they are more chaotic and ultimately do not closely match. This can be best seen with the last style image, the self-portrait by \textit{Alexandr Benois} (Figure~\ref{fig:comparison}, last row).
The results of both methods respect the desired flow of patterns, however, our approach exhibits a more faithful transfer of style features.
This is evident with all selected styles in Figure~\ref{fig:comparison} and is a direct consequence of how the two methods work, as the approach of Wu et al. is not able to force the synthesis of appropriately rotated patterns.

Furthermore, based on our experimentation with their method and the results in their paper, their choice of parameters seems to be suited for images with large uniform regions and with the user guiding the directional field in those regions.
Note that even though the total loss is a combination of losses for style transfer (style loss and content preservation loss) and directional guidance, these losses are not informed about each other. Thus, the method relies on the directional loss creating directional patterns that may or may not be utilized by the style transfer losses to synthesize appropriately rotated style patterns.

\begin{figure*}[htp]
  \centering
  \includegraphics[width=.9\linewidth]{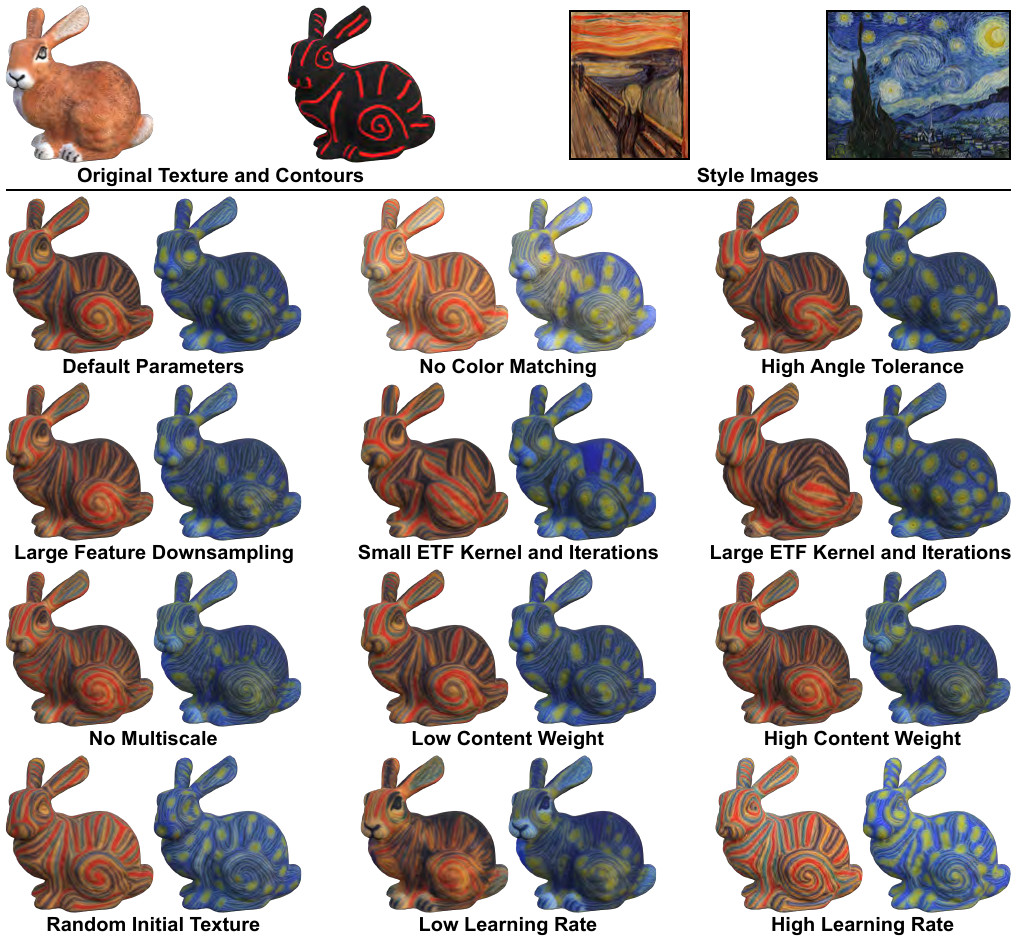}
  \caption{Ablation studies for our approach.}
  \label{fig:ablations}
\end{figure*}

\subsection{Ablation Studies} \label{sec:ablation}

We performed several experiments to evaluate our method and the impact of different parameters on the results, which can be seen in Figure~\ref{fig:ablations}.
We selected the Stanford bunny as the mesh and \textit{The Starry Night} and \textit{The Scream} as the style images, because the impact of certain parameters may not be as strongly visible using only one of the styles.

\noindent \textbf{Color Matching.}
When we do not perform the explicit color transfer step at the beginning, the resulting textures are discolored.
The effect of color matching is shown in Figure~\ref{fig:ablations} (compare "No Color Matching" and "Default Parameters").
Our loss does not explicitly enforce color transfer between the style images and the generated texture---only VGG-16 features of the first few layers.
While these layers contain information about colors, just matching the features does not accurately match color statistics.

\noindent \textbf{Angle Set Granularity.}
During the extraction of style features, we assign them to individual angle sets with a tolerance $\tau$.
Our default choice of $\tau$ is $5^\circ$, in our ablation study, we show the result for $45^\circ$.
The effect of $\tau$ is shown in Figure~\ref{fig:ablations} (see Figure~\ref{fig:ablations}, "High Angle Tolerance" vs. "Default Parameters"). 
Even though $\tau$ is much higher, we still observe that the synthesized patterns approximately correspond to the user-defined directional field, however, the spiral at the hind leg has fewer turns, and the lines at its back are crooked.
The choice of $\tau$ impacts the ability of our method to correctly match the user-defined directional field, with higher granularity, the discretized directional field better corresponds to the continuous directional field, albeit at the cost of more memory and higher optimization time.

\noindent \textbf{Feature Downsampling.}
During the extraction of style features and then during the optimization process, we downsample the extracted features so that all of the feature maps are $\frac{W}{4}\frac{H}{4}$, where $W$ and $H$ are the width and height of the image.
For comparison, we show the result of downsampling by 8---therefore 64 times fewer features---and we observe that, when compared to being downsampled by 4, the brush strokes are less defined and there are noise-like artifacts present (see Figure~\ref{fig:ablations}, "Large Feature Downsampling" vs. "Default Parameters").
The optimization process relies on finding nearest neighbors in feature space---thus, by having fewer of them, fewer patterns and details may be transferred.

\noindent \textbf{Edge Tangent Flow Parameters.}
We use the edge tangent flow to estimate the directions of the patterns in style images and also to compute the directional field of the contours.
Kang et al.'s~\cite{kang2007coherent} algorithm is parametrized by the kernel radius and the number of iterations, both influencing how smooth the resulting field is and how big the patterns must be to be recognized as flowing in a different direction than their surrounding.
By default, we set these parameters to 10 for style images and to 5 for contours.
We present two experiments, in the first one we set the kernel radius and the number of iterations to 1, and in the second one we set them to 20 (see Figure~\ref{fig:ablations}, "Small" vs. "Large ETF Kernel and Iterations").
In both cases, we observe that the patterns do not flow in the desired direction, unlike with the default parameters we use.
While the choice of these parameters is to a certain extent subjective, they need to be large enough to be resistant to noise but small enough to assign meaningful directions to patterns, so that they may be properly recognized as flowing in a certain direction.
Furthermore, we rely on extracted features computed by a convolutional neural network, thus, the features correspond to \textit{overlapping areas} in the images, not just \textit{single pixels}.
For this reason, using a small kernel radius and a few iterations may result in a directional field that is of a higher frequency than the feature maps, making the directional field inaccurate for the extracted features.

\noindent \textbf{Multiscale Style Transfer.}
We optimize the texture twice, once at half of the original resolution and once at the original resolution,  synthesizing patterns at different scales.
When upsampling to start the second stage, we combine the original color-matched unoptimized texture with the generated texture.
We perform three experiments (see Figure~\ref{fig:ablations}, "No Multiscale" vs. "Low" vs. "High Content Weight").
In the first one, we only perform the optimization process at the original resolution ("No Multiscale"), in the second one, we do not blend the optimized texture with the original texture ("Low Content Weight"), and in the third one, we blend with a factor of 0.95, giving more weight to the original texture ("High Content Weight").
In the first experiment, we observe that the generated patterns are smaller, namely the stars when using \textit{The Starry Night} as style.
In the second one, we see that the stylization is stronger and less content is preserved compared to our default choice of the blending weight, which is 0.25.
And in the third one, we observe, that much of the original content is preserved and the stylization is much weaker.
Ultimately, the choice of this parameter is a matter of personal choice and depends on the user's intended outcome.

\noindent \textbf{Randomly Initialized Content Texture.}
Instead of using a texture that contains some kind of meaningful content, e.g., a texture created by an artist, we can also use a texture that was initialized with noise (see Figure~\ref{fig:ablations}, "Random Initial Texture").
In that case, the synthesized patterns are random, their randomness corresponding to the hints of patterns in the random texture that will be amplified during the optimization process while being constrained by the directional guidance and, of course, the choice of the style image.
For this experiment, we used a random uniform noise $[0, 1]$, and we observed that with the original content texture, the mouth and paw areas still clearly resemble a mouth and paws, which is not the case for the result generated with the random texture.
Similarly, in the original texture, the bunny's back contains some darker flecks, which result in darker stylization in those areas, however, when using the random texture, the areas have approximately the same distribution of patterns as the rest of the texture.

\noindent \textbf{Learning Rate.}
The loss we use is based on approaching nearest neighbors in a very high-dimensional feature space.
By modifying the learning rate (see Figure~\ref{fig:ablations}, "Low" vs. "High Learning Rate"), we influence how quickly the optimized features approach their nearest neighbors, and if the learning rate is large enough, they can overshoot and the new nearest neighbors may be different than the original ones as noted by Kolkin et al.~\cite{kolkin2022nnst}.
Thus, by changing the learning rate, we can modulate how much the original content is preserved, although with a higher learning rate, noise-like artifacts start to appear which may be partially compensated for by increasing the weight of the total variation loss. With a smaller learning rate, the original content is more preserved.

\subsection{Performance}

We measured the optimization time of our method on a PC with an NVIDIA GeForce RTX 4080 SUPER GPU with 16 GB of VRAM.
Computing the rotated style features of the style images we use takes approximately 15 seconds.
With the default parameters and when using only a single style image, Style Brush takes approximately 8 minutes to create a stylized texture, of which the precomputation part takes 1 minute and the optimization takes 7 minutes.
When using multiple styles, the optimization takes $n$ times longer for $n$ style images.

\section{Limitations}

By utilizing a loss based on finding nearest neighbors in feature space, we also inherit its limitations.
Namely, the nearest neighbor search has a time complexity of $O(n^2)$, which is relatively slow.
An approximate nearest neighbor search~\cite{jegou2010anns, zhao2020anns} might offer a speed up at a potential loss of stylization quality, though a proper evaluation should be performed first to evaluate its impact.
Furthermore, these nearest neighbor losses generally utilize a simple CNN, e.g., VGG-16, to extract features.
However, more advanced neural approaches for style transfer might utilize different architectures, e.g., transformers, or a completely different way to approach this problem, e.g., by utilizing a diffusion model.
In such cases, assigning directionality to the extracted feature vectors like we do, may not be feasible, and a fundamentally different way to offer directional guidance might be necessary.
Moreover, GPUs of today have a relatively small VRAM.
We rely on extracting and storing style features of a large number of rotated style images.
When using large style images or using many of them, we may have to rely on the slower RAM and then copy the style features to the VRAM on demand, or drop quality by downsampling style features, or increasing the angle tolerance.
Lastly, we utilize the Fibonacci sphere to place cameras around the object, which may cause certain areas to take longer to converge than necessary if they are not frequently seen from the viewpoints.
A detailed analysis of the object or scenes to place cameras could lead to a faster convergence.

\section{Conclusions and Future Work}

We introduced a novel algorithm, Style Brush, for stylizing textured meshes to match the style of given style images while respecting additional directional guidance.
By optimizing the texture with a nearest neighbor-based loss that filters style features on their directionality, we allow the users of our method to guide the stylization process to create high-quality stylized textures in a few minutes.
To our knowledge, we propose the first method that allows this kind of guidance for 3D objects.
In the future, we aim to explore alternative networks for guided style transfer and potentially extend our method to networks that do not easily associate extracted feature vectors with their directional properties.
Finally, one could consider modifying the geometry of the provided mesh to better suit the provided style.

\bibliographystyle{eg-alpha-doi}  
\bibliography{egbibsample}        

\newcommand{\etalchar}[1]{$^{#1}$}
\begin{thebibliography}{\uppercase{KFCO{\etalchar{*}}07}}

\bibitem[AHS{\etalchar{*}}21]{an2021artflowff}
\textsc{An J., Huang S., Song Y., Dou D., Liu W., Luo J.}:
\newblock Artflow: Unbiased image style transfer via reversible neural flows.
\newblock In \emph{Proceedings of the IEEE/CVF Conference on Computer Vision and Pattern Recognition} (2021).

\bibitem[CHH24]{chung2024style}
\textsc{Chung J., Hyun S., Heo J.-P.}:
\newblock Style injection in diffusion: A training-free approach for adapting large-scale diffusion models for style transfer.
\newblock In \emph{Proceedings of the IEEE/CVF International Conference on Computer Vision} (2024).

\bibitem[CS16]{chen2016styletransferfastnn}
\textsc{Chen T.~Q., Schmidt M.}:
\newblock Fast patch-based style transfer of arbitrary style.
\newblock In \emph{arXiv preprint arXiv:1612.04337} (2016).

\bibitem[CW10]{chen2010high}
\textsc{Chen J., Wang B.}:
\newblock High quality solid texture synthesis using position and index histogram matching.
\newblock \emph{The Visual Computer} (2010).

\bibitem[CWNN20]{cao2020styletransferpointclouds}
\textsc{Cao X., Wang W., Nagao K., Nakamura R.}:
\newblock {PSNet: A Style Transfer Network for Point Cloud Stylization on Geometry and Color}.
\newblock In \emph{2020 IEEE Winter Conference on Applications of Computer Vision (WACV)} (2020).

\bibitem[CXG{\etalchar{*}}22]{chen2022tensorf}
\textsc{Chen A., Xu Z., Geiger A., Yu J., Su H.}:
\newblock Tensorf: Tensorial radiance fields.
\newblock In \emph{European Conference on Computer Vision (ECCV)} (2022).

\bibitem[GCLY18]{gu2018arbitrarystyleloss}
\textsc{Gu S., Chen C., Liao J., Yuan L.}:
\newblock Arbitrary style transfer with deep feature reshuffle.
\newblock In \emph{Proceedings of the IEEE conference on computer vision and pattern recognition} (2018).

\bibitem[GEB15a]{gatys2015styletransfer}
\textsc{Gatys L.~A., Ecker A.~S., Bethge M.}:
\newblock A neural algorithm of artistic style.
\newblock \emph{arXiv preprint arXiv:1508.06576} (2015).

\bibitem[GEB15b]{gatys2015texturesynthesis}
\textsc{Gatys L.~A., Ecker A.~S., Bethge M.}:
\newblock Texture synthesis using convolutional neural networks.
\newblock \emph{Advances in neural information processing systems 28} (2015).

\bibitem[GRGH19]{gutierrez2019volumetexture}
\textsc{Gutierrez J., Rabin J., Galerne B., Hurtut T.}:
\newblock {On Demand Solid Texture Synthesis Using Deep {3D} Networks}.
\newblock \emph{Computer Graphics Forum} (2019).

\bibitem[HB17]{huang2017arbitrarytransferff}
\textsc{Huang X., Belongie S.}:
\newblock Arbitrary style transfer in real-time with adaptive instance normalization.
\newblock In \emph{Proceedings of the IEEE international conference on computer vision} (2017).

\bibitem[HJN22]{hollein2022styletransferrender}
\textsc{H{\"o}llein L., Johnson J., Nie{\ss}ner M.}:
\newblock {Stylemesh: Style Transfer for Indoor {3D} Scene Reconstructions}.
\newblock In \emph{Proceedings of the IEEE/CVF Conference on Computer Vision and Pattern Recognition (CVPR)} (2022).

\bibitem[HMR20]{henzler2020neuraltexture}
\textsc{Henzler P., Mitra N.~J., Ritschel T.}:
\newblock {Learning a Neural {3D} Texture Space from {2D} Exemplars}.
\newblock In \emph{The IEEE Conference on Computer Vision and Pattern Recognition (CVPR)} (2020).

\bibitem[IZZE17]{isola2017image}
\textsc{Isola P., Zhu J.-Y., Zhou T., Efros A.~A.}:
\newblock Image-to-image translation with conditional adversarial networks.
\newblock In \emph{Proceedings of the IEEE conference on computer vision and pattern recognition} (2017), pp.~1125--1134.

\bibitem[JBV17]{jetchev2017texturesynthesisgan}
\textsc{Jetchev N., Bergmann U., Vollgraf R.}:
\newblock Texture synthesis with spatial generative adversarial networks.
\newblock \emph{arXiv preprint arXiv:1611.08207} (2017).

\bibitem[JDS10]{jegou2010anns}
\textsc{Jegou H., Douze M., Schmid C.}:
\newblock Product quantization for nearest neighbor search.
\newblock \emph{IEEE transactions on pattern analysis and machine intelligence 33}, 1 (2010), 117--128.

\bibitem[KFCO{\etalchar{*}}07]{kopf2007solid}
\textsc{Kopf J., Fu C.-W., Cohen-Or D., Deussen O., Lischinski D., Wong T.-T.}:
\newblock {Solid Texture Synthesis from {2D} Exemplars}.
\newblock In \emph{ACM SIGGRAPH 2007}. ACM, 2007.

\bibitem[KHR24a]{kovacs2024gstyle}
\textsc{Kov{\'a}cs {\'A}.~S., Hermosilla P., Raidou R.~G.}:
\newblock G-style: Stylized gaussian splatting.
\newblock In \emph{Computer Graphics Forum} (2024), vol.~43, Wiley Online Library, p.~e15259.

\bibitem[KHR24b]{kovacs2024surface}
\textsc{Kov{\'a}cs {\'A}.~S., Hermosilla P., Raidou R.~G.}:
\newblock Surface-aware mesh texture synthesis with pre-trained {2D} cnns.
\newblock In \emph{Computer Graphics Forum (Eurographics)} (2024).

\bibitem[KKLD23]{kerbl3dgaussians}
\textsc{Kerbl B., Kopanas G., Leimk{\"u}hler T., Drettakis G.}:
\newblock {3D} gaussian splatting for real-time radiance field rendering.
\newblock \emph{ACM Transactions on Graphics} (2023).

\bibitem[KKP{\etalchar{*}}22]{kolkin2022nnst}
\textsc{Kolkin N., Kucera M., Paris S., Sykora D., Shechtman E., Shakhnarovich G.}:
\newblock Neural neighbor style transfer.
\newblock \emph{arXiv preprint arXiv:2203.13215} (2022).

\bibitem[KLC07]{kang2007coherent}
\textsc{Kang H., Lee S., Chui C.~K.}:
\newblock Coherent line drawing.
\newblock In \emph{Proceedings of the 5th international symposium on Non-photorealistic animation and rendering} (2007), pp.~43--50.

\bibitem[KSS19]{kolkin2019styleoptimaltransportlossnn}
\textsc{Kolkin N., Salavon J., Shakhnarovich G.}:
\newblock Style transfer by relaxed optimal transport and self-similarity.
\newblock In \emph{Proceedings of the IEEE/CVF conference on computer vision and pattern recognition} (2019).

\bibitem[LW16]{li2016combiningmarkovnn}
\textsc{Li C., Wand M.}:
\newblock Combining markov random fields and convolutional neural networks for image synthesis.
\newblock \emph{IEEE Conference on Computer Vision and Pattern Recognition (CVPR)} (2016).

\bibitem[LYY{\etalchar{*}}17]{liao2017imageanalogylossnn}
\textsc{Liao J., Yao Y., Yuan L., Hua G., Kang S.~B.}:
\newblock Visual attribute transfer through deep image analogy.
\newblock \emph{ACM Transactions on Graphics} (2017).

\bibitem[LZC{\etalchar{*}}23]{liu2023stylerf}
\textsc{Liu K., Zhan F., Chen Y., Zhang J., Yu Y., Saddik A.~E., Lu S., Xing E.}:
\newblock Stylerf: Zero-shot {3D} style transfer of neural radiance fields.
\newblock \emph{Proc. IEEE Conf. on Computer Vision and Pattern Recognition (CVPR)} (2023).

\bibitem[LZL{\etalchar{*}}23]{liu2023instant}
\textsc{Liu R., Zhao E., Liu Z., Feng A., Easley S.~J.}:
\newblock Instant photorealistic style transfer: A lightweight and adaptive approach.
\newblock \emph{arXiv preprint arXiv:2309.10011} (2023).

\bibitem[MPSO18]{mordvintsev2018styletransferrender}
\textsc{Mordvintsev A., Pezzotti N., Schubert L., Olah C.}:
\newblock Differentiable image parameterizations.
\newblock \emph{Distill 3}, 7 (2018), e12.

\bibitem[MST{\etalchar{*}}20]{mildenhall2020nerf}
\textsc{Mildenhall B., Srinivasan P.~P., Tancik M., Barron J.~T., Ramamoorthi R., Ng R.}:
\newblock {NeRF: Representing Scenes as Neural Radiance Fields for View Synthesis}.
\newblock In \emph{European Conference on Computer Vision (ECCV)} (2020).

\bibitem[PLWZ19]{park2019semantic}
\textsc{Park T., Liu M.-Y., Wang T.-C., Zhu J.-Y.}:
\newblock Semantic image synthesis with spatially-adaptive normalization.
\newblock In \emph{Proceedings of the IEEE/CVF conference on computer vision and pattern recognition} (2019), pp.~2337--2346.

\bibitem[RBS{\etalchar{*}}22]{reimann2022controlling}
\textsc{Reimann M., Buchheim B., Semmo A., D{\"o}llner J., Trapp M.}:
\newblock Controlling strokes in fast neural style transfer using content transforms.
\newblock \emph{The Visual Computer 38}, 12 (2022), 4019--4033.

\bibitem[SZ14]{simonyan2014vgg16}
\textsc{Simonyan K., Zisserman A.}:
\newblock Very deep convolutional networks for large-scale image recognition.
\newblock \emph{International Conference on Learning Representations (ICLR)} (2014).

\bibitem[WLZ{\etalchar{*}}18]{wang2018high}
\textsc{Wang T.-C., Liu M.-Y., Zhu J.-Y., Tao A., Kautz J., Catanzaro B.}:
\newblock High-resolution image synthesis and semantic manipulation with conditional gans.
\newblock In \emph{Proceedings of the IEEE conference on computer vision and pattern recognition} (2018), pp.~8798--8807.

\bibitem[WSZL19]{wu2019direction}
\textsc{Wu H., Sun Z., Zhang Y., Li Q.}:
\newblock Direction-aware neural style transfer with texture enhancement.
\newblock \emph{Neurocomputing 370} (2019), 39--55.

\bibitem[WZX23]{wang2023stylediffusion}
\textsc{Wang Z., Zhao L., Xing W.}:
\newblock Stylediffusion: Controllable disentangled style transfer via diffusion models.
\newblock In \emph{Proceedings of the IEEE/CVF International Conference on Computer Vision} (2023).

\bibitem[XCX{\etalchar{*}}24]{xu2024styledyrf}
\textsc{Xu H., Chen W., Xiao F., Sun B., Kang W.}:
\newblock {StyleDyRF: Zero-shot 4D Style Transfer for Dynamic Neural Radiance Fields}.
\newblock \emph{arXiv preprint arXiv:2403.08310} (2024).

\bibitem[ZFLS24]{zhang2024coarf}
\textsc{Zhang D., Fernandez-Labrador C., Schroers C.}:
\newblock {CoARF: Controllable {3D} Artistic Style Transfer for Radiance Fields}.
\newblock \emph{Internation Conference on {3D} Vision ({3D}V)} (2024).

\bibitem[ZGW{\etalchar{*}}22]{zhao2022stsgan}
\textsc{Zhao X., Guo J., Wang L., Li F., Zheng J., Yang B.}:
\newblock {STS-GAN: Can We Synthesize Solid Texture with High Fidelity from Arbitrary Exemplars?}
\newblock \emph{Proceedings of the Thirty-Second International Joint Conference on Artificial Intelligence} (2022).

\bibitem[ZHT{\etalchar{*}}23]{zhang2023diffusionstyletransfer}
\textsc{Zhang Y., Huang N., Tang F., Huang H., Ma C., Dong W., Xu C.}:
\newblock Inversion-based style transfer with diffusion models.
\newblock In \emph{Proceedings of the IEEE/CVF Conference on Computer Vision and Pattern Recognition (CVPR)} (2023).

\bibitem[ZKB{\etalchar{*}}22]{zhang2022arf}
\textsc{Zhang K., Kolkin N., Bi S., Luan F., Xu Z., Shechtman E., Snavely N.}:
\newblock {ARF: Artistic radiance fields}.
\newblock In \emph{European Conference on Computer Vision} (2022).

\bibitem[ZTL20]{zhao2020anns}
\textsc{Zhao W., Tan S., Li P.}:
\newblock Song: Approximate nearest neighbor search on gpu.
\newblock In \emph{2020 IEEE 36th International Conference on Data Engineering (ICDE)} (2020), IEEE, pp.~1033--1044.

\end{thebibliography}

\end{document}